\documentclass[12pt,preprint]{aastex}

\begin{document}

\slugcomment{KSUPT--07/2 \quad June 2007}

\title{Galaxy Cluster Gas Mass Fraction and Hubble Parameter versus Redshift Constraints on Dark Energy}

\author{Lado Samushia\altaffilmark{1,2,3}, Gang Chen\altaffilmark{4,5}, and Bharat Ratra\altaffilmark{1,6}}

\altaffiltext{1}{Department of Physics, Kansas State University, 116 Cardwell Hall, Manhattan, KS 66506}
\altaffiltext{2}{National Abastumani Astrophysical Observatory, 2A Kazbegi Ave, GE-0160 Tbilisi, Georgia}
\altaffiltext{3}{email: lado@phys.ksu.edu}
\altaffiltext{4}{Institute for Astronomy, 2680 Woodlawn Drive, Honolulu, HI 96822}
\altaffiltext{5}{email: gchen@ifa.hawaii.edu}
\altaffiltext{6}{email: ratra@phys.ksu.edu}

\begin{abstract}
Galaxy cluster gas mass fraction versus redshift data \citep{all04} and Hubble parameter versus redshift data \citep{sim05} are used to jointly constrain dark energy models. These constraints favor the Einstein cosmological constant limit of dark energy but do not strongly rule out slowly-evolving dark energy.
\end{abstract}


\section{Introduction}
Type Ia supernova apparent magnitude versus redshift data now favor nonzero dark energy at about four standard deviations \citep[see, e.g.,][]{clo06, ast06, rie07, dav07}. Consistent with this, cosmic microwave background anisotropy measurements indicate that the Universe is spatially flat \citetext{see, e.g., \citealp{pod01, dur03, pag03, spe06}, if the dark energy density is assumed to be a constant, see, e.g., \citealp{wri06, wan07}}, and, in conjunction with the low observed nonrelativistic matter density \citep[see, e.g.,][]{che03b, spe06}, imply that dark energy (DE) accounts for $\sim 70\ \%$ of the Universe's energy budget. 

A number of explanations have been proposed for the DE phenomena. DE might be a cosmological constant  \citep{pee84} or it could be a dynamic scalar field with negative pressure \citep{pee88}.\footnote{Alternatively, it could be that general relativity needs to be modified on very large scales (see, e.g., \citealt{wan07a, mov07, tsu07, eli07}).} For recent dark energy reviews see, e.g., \citet{pee03}, \citet{pad05}, \citet{cop06}, and \citet{nob06}.

Since different DE models make different predictions for the expansion history of the Universe and for the growth of perturbations, DE model parameters can be constrained by using available cosmological observations. Observations such as Type Ia supernova (SNIa) apparent luminosity versus redshift \citep[see, e.g.,][]{nes06, jas06, bar07}; cosmic microwave background anisotropy \citep[see, e.g.,][]{muk03, spe06}; the angular size versus redshift relation for quasars and radio sources \citep[see, e.g.,][]{che03a, pod03, dal06}; strong gravitational lensing by a foreground galaxy or cluster of galaxies \citep[see, e.g.,][]{cha04, alc05a, fed07}; and various large-scale structure measurements \citep[see, e.g.,][]{sel05, teg06, per07}, including baryon acoustic peak measurements \citep[see, e.g.,][]{eis05, wan06, dor07, par07}, and galaxy cluster number counts \citep[see, e.g.,][]{voi05, you05}, may be used to constrain model parameters. 

Since most observables depend on combinations of cosmological parameters rather then on just a single parameter, a single data set can not provide strong constraints. To get around this it is important to consider many different cosmological tests. This allows for consistency checks and might also allow for identification of systematic effects present in a particular data set. Combining data sets with constraints that are orthogonal to each other in parameter space results in significantly tighter constraints.

In this paper we use galaxy cluster gas mass fraction versus redshift data \citetext{\citealt{all04}, also see \citealt{sas96, pen97}} and Hubble parameter versus redshift data \citetext{\citealt{sim05}, also see \citealt{jim02}} to jointly constrain parameters of three different dark energy models. The first model we study is the cosmological constant dominated cold dark matter model ($\Lambda$CDM) with redshift-independent cosmological constant energy density parameter $\Omega_\Lambda$. We also consider the XCDM parametrization of dark energy, where dark energy is taken to be a fluid with an equation of state that relates pressure $p_{\rm x}=\omega_{\rm x}\rho_{\rm x}$ to the energy density $\rho_{\rm x}$, where $\omega_{\rm x}$ is a negative constant (this is only an approximate parametrization of dark energy). Thirdly, we consider a slowly-rolling dark energy scalar field model ($\phi$CDM) in which the scalar field $\phi$ has potential energy density $V(\phi) \propto \phi^{-\alpha}$, where $\alpha$ is a nonnegative constant \citep{pee88, rat88}. For the $\phi$CDM and XCDM cases we only consider spatially-flat spacetimes, while in the $\Lambda$CDM model spatial curvature is allowed to be nonzero.  XCDM and $\phi$CDM reduce to the  time-independent dark energy $\Lambda$CDM model when  $\omega_{\rm x}=-1$ and $\alpha=0$, respectively. In this paper we jointly analyze both data sets and derive constraints on the nonrelativistic matter density parameter $\Omega_{\rm m}$ and a parameter $p$ that describes the DE. The parameter $p$ is $\Omega_\Lambda$ for $\Lambda$CDM, $\omega_{\rm x}$ for XCDM, and $\alpha$ for $\phi$CDM.

The galaxy cluster gas mass fraction versus redshift data has been used to constrain parameters of the $\Lambda$CDM, XCDM and $\phi$CDM models \citep{all04, che04}. These data provide tight constraints on $\Omega_{\rm m}$. \citet{rap05} used the galaxy cluster data in combination with CMB anisotropy and SNIa measurements to constrain dark energy evolution. For the XCDM model, assuming a time-independent equation of state, they set  tight limits, $\omega_{\rm x}=-1.05^{+ 0.10}_{-0.12}$, while more generally they found no significant evidence for evolution in the dark energy equation of state.  \citet{wil06} used these data in combination with SNIa data and found that the joint constraints were significantly tighter then those derived from either data set alone; the combined analysis favored the $\Lambda$CDM model but did not strongly rule out slowly-evolving dark energy. \citet{alc05b} used the galaxy cluster data and SNIa data (along with priors on the Hubble parameter and the baryonic matter density) to jointly constrain brane world models. This data set has been used in conjunction with Fanaroff-Riley type IIb radio galaxy angular size distance measurements to put an upper limit on the amplitude of non-Riemannian terms during the late stages of the Universe's evolution \citep{pue05}. Galaxy cluster gas mass fraction data have also been used to constrain other dark energy models \citep[see, e.g.,][]{cha06, zha06}.

The $H(z)$ data were used by \citet{sam06} to constrain cosmological parameters in the $\Lambda$CDM, XCDM and $\phi$CDM models, but a computational error was made when cosmological parameter confidence contours were calculated. \citet{sen07} used these data to constrain the evolution of an arbitrary dark energy component that satisfies the weak energy condition, in spatially-flat models. The $H(z)$ data set has also been used to constrain a number of interacting dark energy models \citep{wei07a, wei07b, zha07}. In combination with CMB anisotropy measurements and SNIa data it has been used to constrain the Chaplygin gas model \citep{wu07} as well as  cosmological models motivated by higher dimensional theories \citep{laz07}. 

In this paper we present corrected cosmological parameter constraints for the $H(z)$ data. We also provide joint constraints on the $\Lambda$CDM, XCDM, and $\phi$CDM models from the $H(z)$ and galaxy cluster gas mass fraction versus redshift data. In Sec.\ 2 we outline our computational method. Results are presented and discussed in Sec.\ 3.

\section{Computation}
We use the \citet{all04} measurements of gas mass fractions for 26 relaxed rich clusters in the redshift range $0.08<z<0.89$. The cluster baryon mass is dominated by the gas. In relaxed rich clusters the baryon fraction should be independent of redshift. The cluster baryon fraction value depends on the angular diameter distance, so the correct cosmological parameter values place clusters at the right angular diameter distance to ensure the redshift independence of the cluster baryon fraction. We follow \citet{che04} and compute the two dimensional likelihood function $L^G(\Omega_{\rm m}, p)$ for each of the three DE models. When computing $L^G(\Omega_{\rm m}, p)$ we marginalize over the Gaussian uncertainties in the bias factor $b$, in the Hubble constant $h$ (in units of $100\ {\rm km\, s}^{-1}{\rm Mpc}^{-1}$), and in the baryonic matter density parameter $\Omega_{\rm b}$. Following \citet{all04}, we use $b=0.824\pm0.089$ (one standard deviation error) for the bias factor. To reflect the range of uncertainties, we use two sets of values for $h$ and $\Omega_{\rm b} h^2$. One set is $\Omega_{\rm b} h^2=0.014\pm0.004$ (one standard deviation error, \citealt{pee03}) and $h=0.68\pm0.04$ (one standard deviation error, \citealt{got01, che03}). The other is from the WMAP three-year data, $\Omega_{\rm b}h^2=0.0228\pm0.0007$ and $h=0.73\pm0.03$ (one standard deviation errors, \citealt{spe06}).

The second data set we use are the nine \citet{sim05} measurements of the Hubble parameter in the  redshift range $0.09<z<1.75$. Following \citet{sam06} we compute a two dimensional likelihood function $L^H(\Omega_{\rm m}, p)$ for each DE model. $H(z)$ is not sensitive to the bias factor or baryonic matter density, but we still have to account for uncertainties in the  Hubble constant. For the Hubble constant prior probability distribution function we use the same set of values as in the previous paragraph.

To derive joint constraints, for each DE model we define the joint likelihood function $L(\Omega_{\rm m}, p)=L^G(\Omega_{\rm m}, p)L^H(\Omega_{\rm m}, p)$. From the joint likelihood function we compute 1, 2, and 3 $\sigma$ confidence contours, as the contours that enclose 68, 95, and 99 $\%$ of the total probability.

\section{Discussion and Conclusion}

\begin{figure}
\epsscale{1.0}
\includegraphics[angle=270]{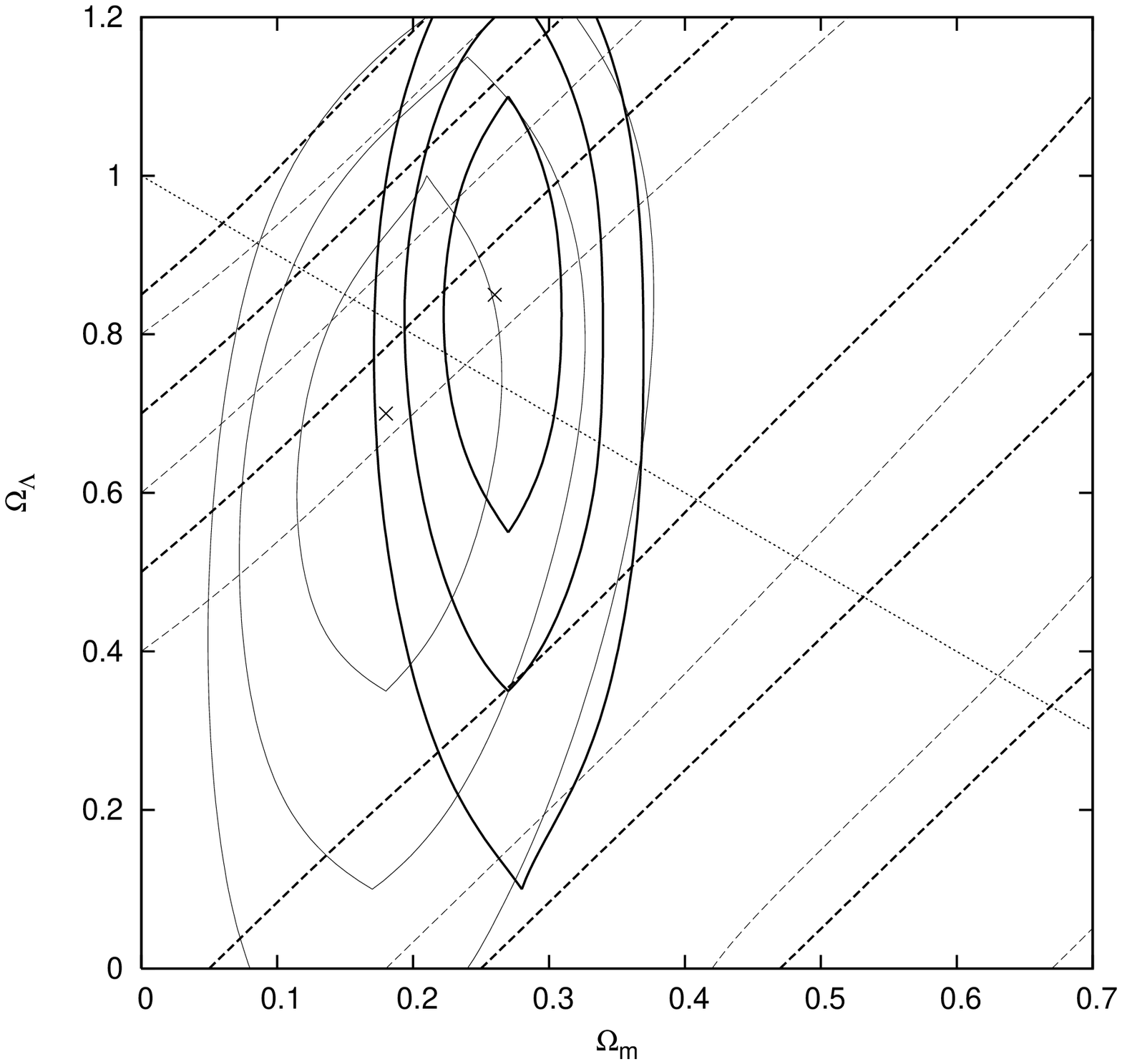}
\caption{1, 2, and 3 $\sigma$ confidence level contours for the $\Lambda$CDM model. Dashed lines denote constraints from Hubble parameter versus redshift data, while solid lines show the joint constraints (the crosses indicate the maximum likelihood points). The diagonal dotted line corresponds to spatially-flat $\Lambda$CDM models. Thick lines correspond to the $h=0.73 \pm 0.03$ and $\Omega_{\rm b} h^2=0.022\pm 0.0007$ priors (maximum likelihood is at $\Omega_{\rm m}=0.26$ and $\Omega_\Lambda=0.85$), while thin lines are for $h=0.68 \pm 0.04$ and $\Omega_{\rm b} h^2=0.014\pm 0.04$ (maximum likelihood is at $\Omega_{\rm m}=0.18$ and $\Omega_\Lambda=0.70$).}
\label{lcdm}
\end{figure}

\begin{figure}
\epsscale{1.0}
\includegraphics[angle=270]{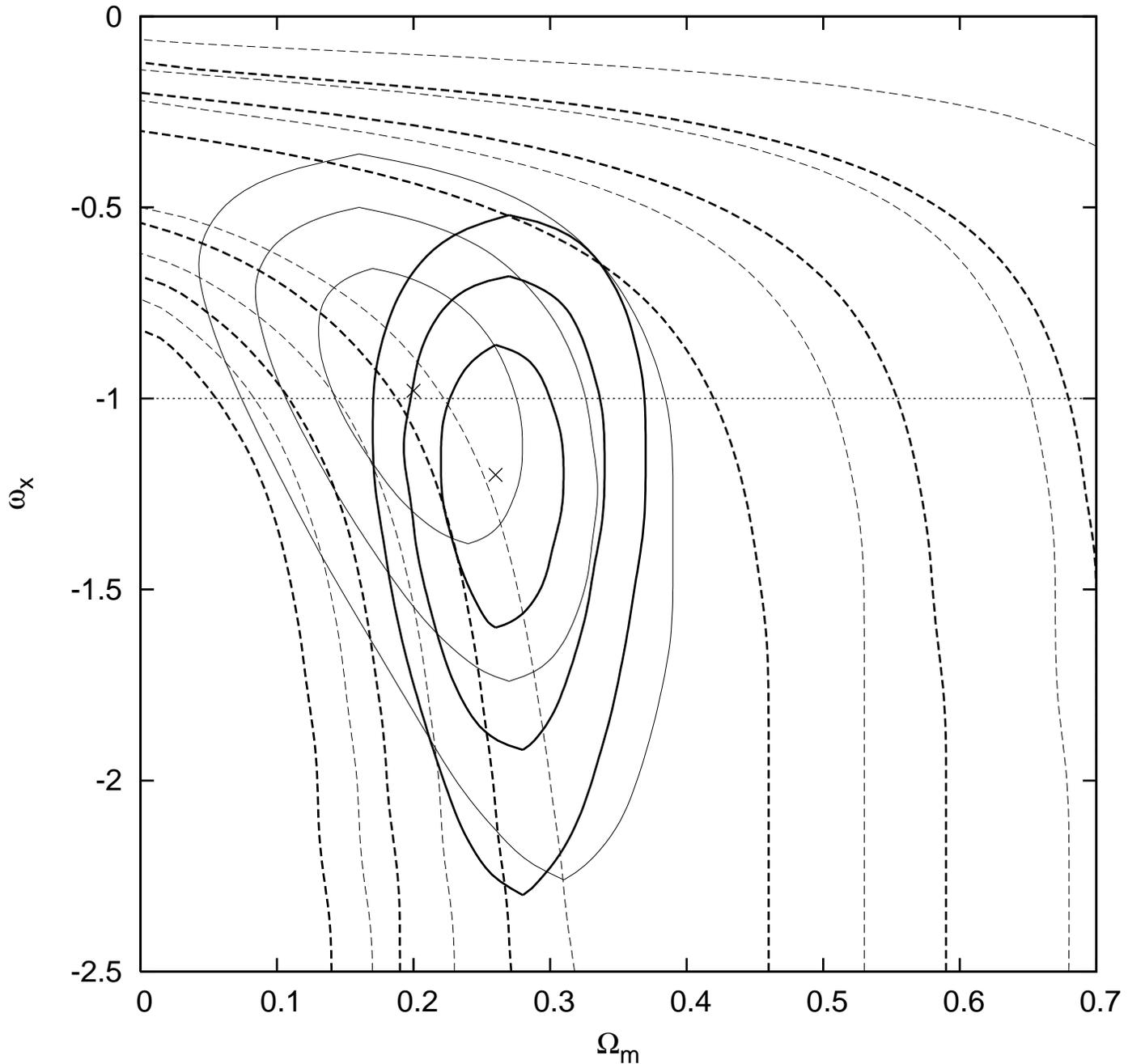}
\caption{1, 2, and 3 $\sigma$ confidence level contours for the XCDM parametrization. Dashed lines denote constraints from Hubble parameter versus redshift data, while solid lines show the joint constraints (the crosses indicate the maximum likelihood points). The dotted horizontal line corresponds to spatially-flat $\Lambda$CDM models. Thick lines correspond to the $h=0.73 \pm 0.03$ and $\Omega_{\rm b} h^2=0.022\pm 0.0007$ priors (maximum likelihood is at $\Omega_{\rm m}=0.26$ and $\omega_{\rm x}=-1.2$), while thin lines are for $h=0.68 \pm 0.04$ and $\Omega_{\rm b} h^2=0.014\pm 0.04$ (maximum likelihood is at $\Omega_{\rm m}=0.20$ and $\omega_{\rm x}=-0.98$).}
\label{xcdm}
\end{figure}

\begin{figure}
\epsscale{1.0}
\includegraphics[angle=270]{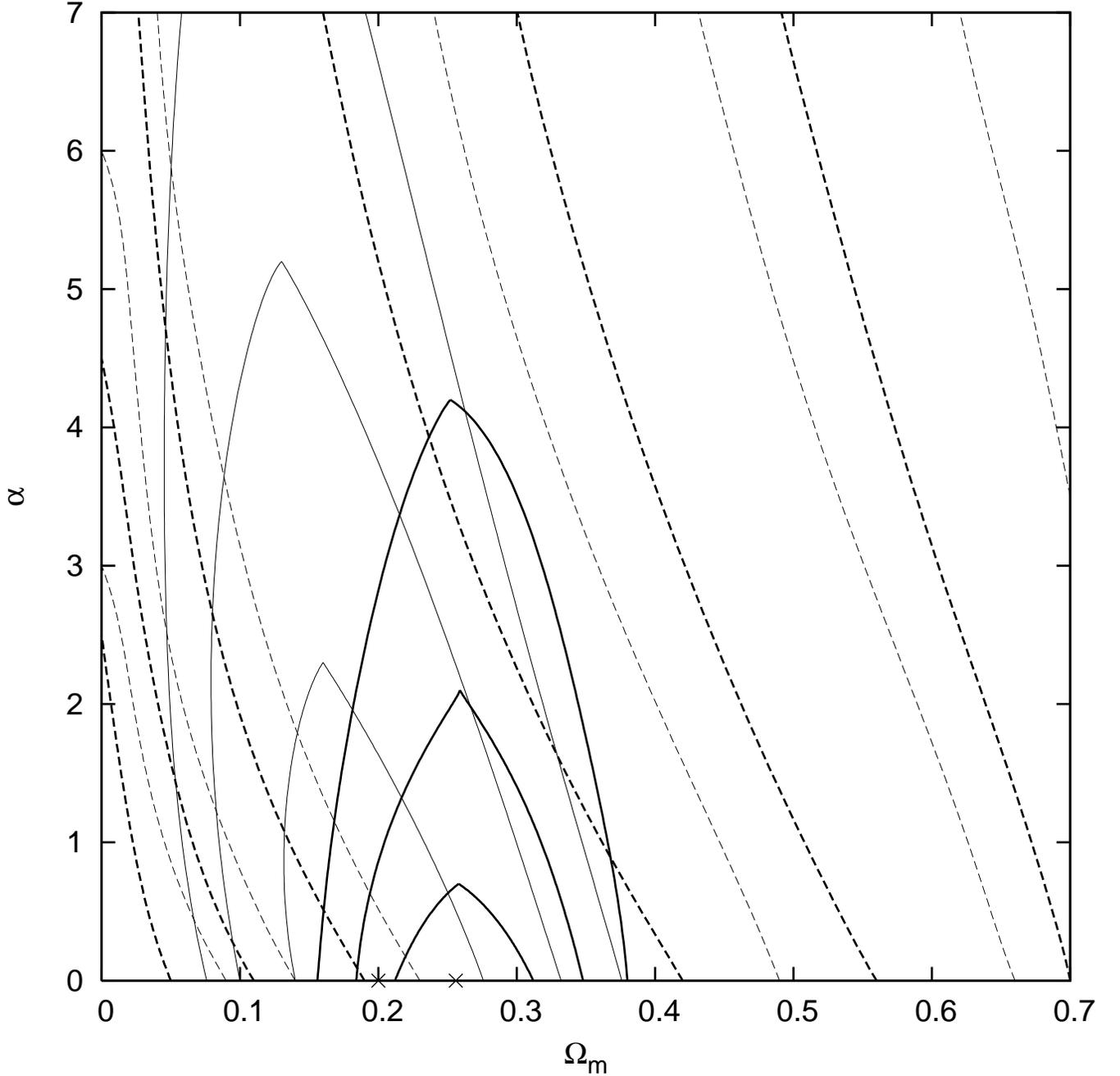}

\caption{1, 2, and 3 $\sigma$ confidence level contours for the $\phi$CDM model. Dashed lines denote constraints from Hubble parameter versus redshift data, while solid lines show the joint constraints (the crosses on the horizontal axis indicate the maximum likelihood points). The horizontal $\alpha=0$ axis corresponds to spatially-flat $\Lambda$CDM models. Thick lines correspond to the $h=0.73 \pm 0.03$ and $\Omega_{\rm b} h^2=0.022\pm 0.0007$ priors (maximum likelihood is at $\Omega_{\rm m}=0.26$ and $\alpha=0$), while thin lines are for $h=0.68 \pm 0.04$ and $\Omega_{\rm b} h^2=0.014\pm 0.04$ (maximum likelihood is at $\Omega_{\rm m}=0.20$ and $\alpha=0$).}
\label{phycdm}
\end{figure}

Figures \ref{lcdm} to \ref{phycdm} show cosmological parameter confidence contours for the $\Lambda$CDM, XCDM and $\phi$CDM models for the two sets of $\Omega_{\rm b} h^2$ and $h$ priors. 

Figure \ref{lcdm} shows constraints on the $\Lambda$CDM model. The galaxy cluster gas mass fraction data place a good constraint on $\Omega_{\rm m}$ ($< 0.35$ at 3 $\sigma$), while the $H(z)$ data constrain a linear combination of $\Omega_{\rm m}$ and $\Omega_\Lambda$. The joint likelihood functions peak near spatially-flat models.

Figure \ref{xcdm} shows the constraints for the XCDM parametrization. The joint constraints favor the region of parameter space near the $\omega_{\rm x}=-1$ line which corresponds to spatially-flat $\Lambda$CDM models.

Figure \ref{phycdm} is for the $\phi$CDM model. The joint likelihoods peak on the $\alpha=0$ line which corresponds to the spatially-flat $\Lambda$CDM model. However, values of $\alpha$ as high as 4 or 5 are allowed at 3 $\sigma$.

The galaxy cluster gas mass fraction data is more restrictive than the $H(z)$ data. When they are combined the $H(z)$ data shifts the constraints to slightly higher values of $\Omega_{\rm m}$ than for the galaxy cluster gas mass fraction data set alone. A spatially-flat cosmological model with a cosmological constant term with $\Omega_\Lambda \simeq 0.7$ is a good fit to the joint data in all six cases considered here. This is consistent with results based on other measurements, see, e.g., \citet{rap05}, \citet{wil06}, and \citet{dav07}.

Hubble parameter versus redshift data is expected to increase by an order of magnitude in the next few years. In combination with new galaxy cluster gas mass fraction, SNIa, and CMB measurements, this will significantly better constrain dark energy models.

We thank R.Lazkoz for helpful discussions. We acknowledge support from DOE grant DE-FG03-99EP41093, INTAS grant 061000017-9258 and NASA ATP grant NAG5-12101.

\begin{thebibliography}{}

\bibitem[Alcaniz et al.(2005)]{alc05a}
  Alcaniz, J. S., Dev, A., \& Jain. D. 2005, \apj, 627, 26

\bibitem[Alcaniz \& Zhu(2005)]{alc05b}
  Alcaniz, J. S., \& Zhu, Z.-H. 2005, \prd, 71, 3153

\bibitem[Allen et al.(2004)]{all04}
  Allen, S. W., et al. 2002, \mnras, 353, 457

\bibitem[Astier et al.(2006)]{ast06}
  Astier, P., et al. 2006. \aap, 447, 31 

\bibitem[Barger et al.(2007)]{bar07}
  Barger, V., Gao, Y., \& Marfatia, D. 2007, Phys. Lett. B, 648, 127

\bibitem[Copeland et al.(2006)]{cop06}
  Copeland, E. J., Sami, M., \& Tsujikawa, S. 2006, Int. J. Mod. Phys.D, 15, 1753

\bibitem[Chae et al.(2004)]{cha04}
  Chae, K.-H., Chen, G., Ratra, B., \& Lee, D. W. 2004, \apj, 607, 71

\bibitem[Chang et al.(2006)]{cha06}
  Chang, Z., et al. 2006, Phys. Lett. B, 633, 14

\bibitem[Chen et al.(2003)]{che03}
  Chen, G., Gott, J. R., \& Ratra, B. 2003, \pasp, 115, 1269

\bibitem[Chen \& Ratra(2003a)]{che03a}
  Chen, G., \& Ratra, B. 2003a, \apj, 582, 586

\bibitem[Chen \& Ratra(2003b)]{che03b}
  Chen, G., \& Ratra, B. 2003b, \pasp, 115, 1143

\bibitem[Chen \& Ratra(2004)]{che04}
  Chen, G., \& Ratra, B. 2004, \apj, 612, 1

\bibitem[Clocchiatti et al.(2006)]{clo06}
  Clocchiatti, A., et al. 2006, \apj, 642, 1

\bibitem[Daly \& Djorgovski(2006)]{dal06}
  Daly, R. A., \& Djorgovski, S. G. 2006, arXiv:astro-ph/0609791

\bibitem[Davis et al.(2007)]{dav07}
  Davis, T. M., et al. 2007, arXiv:astro-ph/0701510

\bibitem[Doran et al.(2007)]{dor07}
  Doran, M., Stern, S., \& Thommes, E. 2007, \jcap, 0704 , 015

\bibitem[Durrer et al.(2003)]{dur03}
  Durrer, R., Novosyadlyj, B., \& Apunevych, S. 2003, \apj, 583, 33

\bibitem[Eisenstein et al.(2005)]{eis05}
  Eisenstein, D. J., et al. 2005, \apj, 633, 560

\bibitem[Elizalde et al.(2007)]{eli07}
  Elizalde, E., et al. 2007, arXiv:0705.1211

\bibitem[Fedeli \& Bartelmann(2007)]{fed07}
  Fedeli, C., \& Bartelmann, M. 2007, \aap, 461, 49

\bibitem[Gott et al.(2001)]{got01}
  Gott, J. R., Vogeley, M. S., Podariu, S., \& Ratra, B. 2001, \apj, 549, 1

\bibitem[Jassal et al.(2006)]{jas06}
  Jassal, H. K., Bagla, J. S., \& Padmanabhan, T. 2006, arXiv:astro-ph/0601389

\bibitem[Jimenez \& Loeb(2002)]{jim02}
  Jimenez, R., \& Loeb, A. 2002, \apj, 573, 37

\bibitem[Lazkoz \& Majerotto(2007)]{laz07}
  Lazkoz, R., \& Majerotto, E. 2007, arXiv:0704.2606

\bibitem[Movahed et al.(2007)]{mov07}
  Movahed, M. S., Baghram, S, \& Rahvar, S. 2007, arXiv:0705.0889

\bibitem[Mukherjee et al.(2003)]{muk03}
  Mukherjee, P, et al. 2003, Int. J. Mod. Phys. A, 18, 4933

\bibitem[Nesseris \& Perivolaropoulos(2006)]{nes06}
  Nesseris. S., \& Perivolaropoulos, L. 2006, \prd, 75, 023517

\bibitem[Nobbenhuis(2006)]{nob06}
  Nobbenhuis, S. 2006, arXiv:gr-qc/0609011

\bibitem[Padmanabhan(2005)]{pad05}
  Padmanabhan, T. 2005, Curr. Sci., 88, 1057

\bibitem[Page et al.(2003)]{pag03}
  Page, L., et al. 2003, \apjs, 148, 223

\bibitem[Parkinson et al.(2007)]{par07}
 Parkinson, D., et al. 2007, \mnras, 377, 185 

\bibitem[Peebles(1984)]{pee84}
  Peebles. P. J. E. 1984, \apj, 284, 439

\bibitem[Peebles \& Ratra(1988)]{pee88}
  Peebles, P. J. E., \& Ratra, B. 1998, \apj, 325, 17

\bibitem[Peebles \& Ratra(2003)]{pee03}
  Peebles, P. J. E., \& Ratra, B. 2003, Rev. Mod. Phys., 75 , 559

\bibitem[Pen(1997)]{pen97}
  Pen, U. L. 1997, \na, 2, 309

\bibitem[Percival et al.(2007)]{per07}
  Percival, W. J., et al. 2007, \apj, 657, 645

\bibitem[Podariu et al.(2003)]{pod03}
  Podariu, S., Daly, R. A., Mory, M. P., \& Ratra, B. 2003, \apj, 584, 577

\bibitem[Podariu et al.(2001)]{pod01}
  Podariu, S., et al. 2001, \apj, 559 , 9

\bibitem[Puetzfeld et al.(2005)]{pue05}
  Puetzfeld, D., Pohl, M., \& Zhu, Z.-H.  2005, \apj, 619, 657

\bibitem[Rapetti et al.(2005)]{rap05}
  Rapetti, D., Allen, S., \& Weller, J. 2005, \mnras, 360, 555

\bibitem[Ratra \& Peebles(1988)]{rat88}
  Ratra, B,. \& Peebles, P. J. E.  1988, \prd, 37, 3406

\bibitem[Riess et al.(2007)]{rie07}
  Riess, A. G., et al. 2007, \apj, 659, 98

\bibitem[Samushia \& Ratra(2006)]{sam06}
  Samushia, L., \& Ratra, B. 2006, \apj, 650, 5

\bibitem[Sasaki(1996)]{sas96}
  Sasaki, S. 1996, \pasj, 48, 119

\bibitem[Seljak et al.(2005)]{sel05}
  Seljak, U., et al. 2005, \prd, 71, 103515 

\bibitem[Sen \& Scherrer(2007)]{sen07}
  Sen, A. A., \& Scherrer, R. J. 2007, arXiv:astro-ph/0703416

\bibitem[Simon et al.(2005)]{sim05}
  Simon, J., Verde, I., \& Jimenez, R. 2005, \prd, 71, 123001

\bibitem[Spergel et al.(2003)]{spe03}
  Spergel, D. N., et al. 2003, \apjs, 148, 175

\bibitem[Spergel et al.(2007)]{spe06}
  Spergel, D. N., et al. 2007, arXiv:astro-ph/0603449

\bibitem[Tegmark et al.(2006)]{teg06}
  Tegmark, M., et al. 2006, \prd, 74, 123507

\bibitem[Tsujikawa(2007)]{tsu07}
  Tsujikawa, S. 2007, arXiv:0705.1032

\bibitem[Voit(2005)]{voi05}
  Voit, G. M. 2005, Rev. Mod. Phys., 77, 207

\bibitem[Wang et al.(2007)]{wan07a}
  Wang, S., Hui, L., May, M., \& Haiman, Z. 2007, arXiv:0705.0165 

\bibitem[Wang(2006)]{wan06}
  Wang, Y. 2006, \apj, 647, 1

\bibitem[Wang \& Mukherjee(2007)]{wan07}
  Wang, Y., \& Mukherjee, P. 2007, arXiv:astro-ph/0703780

\bibitem[Wei \& Zhang(2007a)]{wei07a}
  Wei, H., \& Zhang, S. N. 2007a, Phys. Lett. B, 644, 7 

\bibitem[Wei \& Zhang(2007b)]{wei07b}
  Wei. H., \& Zhang, S. N. 2007b, arXiv:0704.3330

\bibitem[Wilson et al.(2006)]{wil06}
  Wilson, K., Chen. G., \& Ratra, B. 2007, Mod. Phys. Lett. A, 21, 2197

\bibitem[Wright(2006)]{wri06}
  Wright, E. L. 2006, arXiv:astro-ph/0603750

\bibitem[Wu \& Yu(2007)]{wu07}
  Wu, P., \& Yu, H., 2007 Phys. Lett. B, 644, 16

\bibitem[Younger et al.(2005)]{you05}
  Younger, J. D., Bahcall, N. A., \& Bode, P. 2005, \apj, 622, 1

\bibitem[Zhang \& Zhu(2007)]{zha07}
  Zhang, H., \& Zhu, Z.-H. 2007, arXiv:astro-ph/0703245

\bibitem[Zhao et al.(2006)]{zha06}
  Zhao, G. D., Xia, J.-Q., Feng, B., \& Zhang, X. 2006, astro-ph/0603621

\end{thebibliography}
\end{document}